# Tweezers controlled resonator

Samuel Kaminski[1], Leopoldo L. Martin[1] and Tal Carmon[1]

[1]Technion- Israel Institue of Technology, 3200003 Haifa, Israel
[*]tcarmon@technion.ac.il

**Abstract:** We experimentally demonstrate trapping a μdroplet with an optical tweezer and then enabling it as a microresonator by bringing it close to a tapered fiber coupler. Our tweezers facilitated the tuning of the coupling from the under-coupled to the critically coupled regime with an optical Q of 12 million and microresonator size at the 85 μm scale.

**References and links**

1. S. C. Kuo, and M. P. Sheetz, "Force of single kinesin molecules measured with optical tweezers," Science **260**, 232-234 (1993).
2. S. M. Block, L. S. Goldstein, and B. J. Schnapp, "Bead movement by single kinesin molecules studied with optical tweezers,"  (1990).
3. K. C. Neuman, and A. Nagy, "Single-molecule force spectroscopy: optical tweezers, magnetic tweezers and atomic force microscopy," Nature methods **5**, 491-505 (2008).
4. D. G. Grier, "A revolution in optical manipulation," Nature **424**, 810-816 (2003).
5. J. E. Curtis, B. A. Koss, and D. G. Grier, "Dynamic holographic optical tweezers," Optics Communications **207**, 169-175 (2002).
6. A. Terray, J. Oakey, and D. W. Marr, "Fabrication of linear colloidal structures for microfluidic applications," Applied Physics Letters **81**, 1555-1557 (2002).
7. L. Paterson, M. MacDonald, J. Arlt, W. Sibbett, P. Bryant, and K. Dholakia, "Controlled rotation of optically trapped microscopic particles," Science **292**, 912-914 (2001).
8. M. Reicherter, T. Haist, E. Wagemann, and H. Tiziani, "Optical particle trapping with computer-generated holograms written on a liquid-crystal display," Optics letters **24**, 608-610 (1999).
9. J. P. Pantina, and E. M. Furst, "Directed assembly and rupture mechanics of colloidal aggregates," Langmuir **20**, 3940-3946 (2004).
10. A. D. Ward, M. G. Berry, C. D. Mellor, and C. D. Bain, "Optical sculpture: controlled deformation of emulsion droplets with ultralow interfacial tensions using optical tweezers," Chemical communications, 4515-4517 (2006).
11. A. Terray, J. Oakey, and D. W. Marr, "Microfluidic control using colloidal devices," Science **296**, 1841-1844 (2002).
12. P. Galajda, and P. Ormos, "Complex micromachines produced and driven by light," Applied Physics Letters **78**, 249-251 (2001).
13. A. Ashkin, and J. Dziedzic, "Observation of resonances in the radiation pressure on dielectric spheres," Physical Review Letters **38**, 1351 (1977).
14. H.-M. Tzeng, K. F. Wall, M. Long, and R. Chang, "Laser emission from individual droplets at wavelengths corresponding to morphology-dependent resonances," Optics letters **9**, 499-501 (1984).
15. M. Hossein-Zadeh, and K. J. Vahala, "Fiber-taper coupling to Whispering-Gallery modes of fluidic resonators embedded in a liquid medium," Optics express **14**, 10800-10810 (2006).
16. A. Jonáš, Y. Karadag, M. Mestre, and A. Kiraz, "Probing of ultrahigh optical Q-factors of individual liquid microdroplets on superhydrophobic surfaces using tapered optical fiber waveguides," JOSA B **29**, 3240-3247 (2012).
17. U. Levy, K. Campbell, A. Groisman, S. Mookherjea, and Y. Fainman, "On-chip microfluidic tuning of an optical microring resonator," Applied physics letters **88**, 111107 (2006).
18. J. Knight, G. Cheung, F. Jacques, and T. Birks, "Phase-matched excitation of whispering-gallery-mode resonances by a fiber taper," Optics letters **22**, 1129-1131 (1997).
19. M. Cai, O. Painter, and K. J. Vahala, "Observation of critical coupling in a fiber taper to a silica-microsphere whispering-gallery mode system," Physical Review Letters **85**, 74 (2000).
20. S. Spillane, T. Kippenberg, O. Painter, and K. Vahala, "Ideality in a fiber-taper-coupled microresonator system for application to cavity quantum electrodynamics," Physical Review Letters **91**, 043902 (2003).
21. S. Maayani, L. L. Martin, and T. Carmon, "Optical binding in white light," Optics Letters **40**, 1818-1821 (2015).

22.	T. Carmon, L. Yang, and K. Vahala, "Dynamical thermal behavior and thermal self-stability of microcavities," Optics Express **12**, 4742-4750 (2004).
23.	T. Kippenberg, S. Spillane, and K. Vahala, "Kerr-nonlinearity optical parametric oscillation in an ultrahigh-Q toroid microcavity," Physical Review Letters **93**, 083904 (2004).
24.	T. Carmon, and K. J. Vahala, "Visible continuous emission from a silica microphotonic device by third-harmonic generation," Nature Physics **3**, 430-435 (2007).
25.	I. S. Grudinin, A. B. Matsko, and L. Maleki, "Brillouin lasing with a CaF 2 whispering gallery mode resonator," Physical review letters **102**, 043902 (2009).
26.	M. Tomes, and T. Carmon, "Photonic micro-electromechanical systems vibrating at X-band (11-GHz) rates," Physical review letters **102**, 113601 (2009).
27.	G. Bahl, J. Zehnpfennig, M. Tomes, and T. Carmon, "Stimulated optomechanical excitation of surface acoustic waves in a microdevice," Nature communications **2**, 403 (2011).
28.	A. Matsko, A. Savchenkov, V. Ilchenko, D. Seidel, and L. Maleki, "Optomechanics with surface-acoustic-wave whispering-gallery modes," Physical Review Letters **103**, 257403 (2009).
29.	M. L. Gorodetsky, and V. S. Ilchenko, "Optical microsphere resonators: optimal coupling to high-<i>Q</i> whispering-gallery modes," JOSA B **16**, 147-154 (1999).

## 1. Introduction

Optical traps serve in the most sensitive biological-force measurements[1-3] as well as in chemistry and physics research [4]. As optical tweezers can trap almost perfectly spherical droplets while precisely controlling their position, it is natural to check if we can activate tweezed droplets as optical microresonators. By doing so, one can benefit from modern tweezing techniques such as dynamical holograms[5] that simoultanesly manipulate multiple particles [6-9]. Further, traps were shown to sculpt microdroplets into elliptical, triangular and rectangular shapes [10] which can serve in deforming the resonator while light is inside. Additionally pumps and valves [11] were shaped by trapping several adjacent spheres. Such adjacent spheres[11] can serve as cascaded resonators that one can position while controlling each resonator shape and postions. Also, the optical tweezer can be used for multiphotonic photopolymerization of complex a solid structure [12] which was used as a turbine but can serve as an optical component in our regard. These dynamical holograms can uniquely serve in deforming the resonator while light is inside. Interestingly, one of the first experiments in microcavities was done using an optically trapped droplet resonator [13]. In this experiment, Ashkin was tuning the frequency of a laser that was trapping a μdrop against gravity. The upward motion of the drop indicated that the laser frequency was near resonance. Our long-term vision includes optical circuits where a multi-minima optical trap shapes and positions multiple resonators. Being practical, we start here with modestly proving this concept by activating one μdrop as a resonator, and using an optical trap to hold and position it next to a tapered-fiber coupler.

As one can see in figure 1a, our experimental setup consists of optical tweezers (green) that drags a droplet of high index material (yellow) and controls its distance from a nearby tapered fiber (red). Our setup is different from the current state of the art in nanopositioning of optical devices, which uses expensive and cumbersome positioners with an inch-scale footprint. Such nanopositioners are non-scalable to control multiple microresonators. On the contrary, optical tweezers were shown to control many devices with a single beam of light that is controlled by a spatial light modulator (SLM). Droplets were used as optical resonators [13-17] even while optically trapped [13] and here we also control coupling of a tweezed drop resonator to a fiber coupler by moving the optical trap upon need.

## 2. Experimental setup

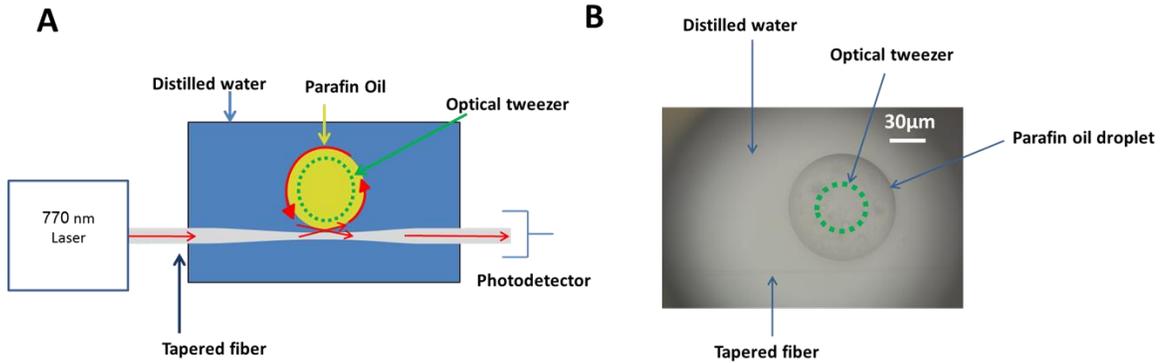

Fig .1. **Experimental setup** (A) where light is coupled from a tapered fiber to circumferentially circulate in a μdrop resonator. (B) A micrograph of our experiment. The dashed green circle represents the location of the optical tweezer. The focuses in the micrograph were optimized separately (for the taper and for the drop).

The optical whispering gallery mode resonators are made by blending together in an ultrasonic bath for half an hour, paraffin oil (0.2ml), distilled water(1ml) and an emulsifier (0.1 ml of "Tween 40"®). An aliquot form of the emulsion was mixed with DI water and injected into a 125 μm deep flow cell that contains a single mode tapered fiber [18-20]. A 532 nm laser with a downward beam that was focused using a GRIN lens (N.A.= 0.46, f=2mm) captures a single droplet on the upper surface of the flow cell, while an upward microscope was used to bottom-view the flow cell. The effective trapping power ranges from 5 to 20 mW using a typical working distance of 7 mm. The optical trap with a relatively low 0.13 NA produces a sufficient force in the direction of the light beam[21] and compensates for the difference in density of both liquids. At the same time, the gradient force that the optical trap applies in the horizontal direction (that is, transverse to beam propagation) serves in dragging the drop to the position where it is nescesary to couple. We trap a drop and use the optical trap to move it towards the tapered fiber until enabling evanescent coupling [18-20] to the droplet [15, 16]. The phase matching that enables efficient coupling is done by tapering the fiber until the speed of light (at waist while in water) is equal to the speed of light in the μdrop resonator. Tuning the laser wavelength through drop's resonanance reveals a dip in the transmission through the fiber with a linewidht inversely proportional to the resonator optical quality factor (Q).

## 3. Experimental results and discussion.

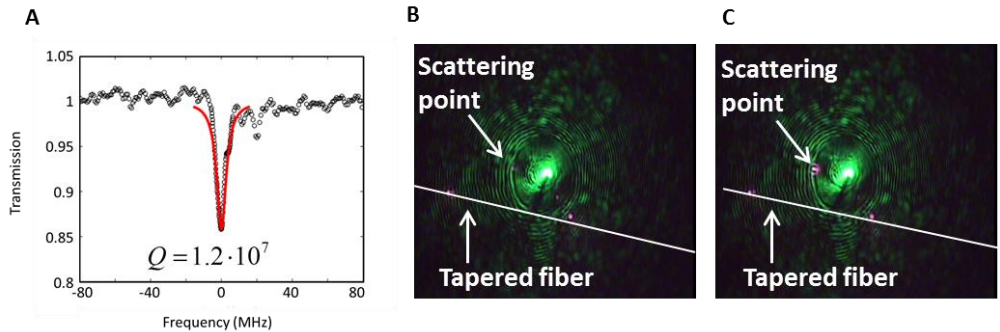

Fig .2. **Experimental results: The optical quality factor,** Q, is experimentally measured by (A) trapping a microsphere and bringing it to the evanescent field of the tapered fiber. When in the undercoupled regime, a

Lorenzian drop in transmission is monitored to fit a 12 million Q. While at resonance, the mode is seen via residual scatterings (C) that are not seen while off resonance (B).

A high quality factor whispering gallery mode resonance in oil-droplet resonator is shown in Fig 2 A, where a Lorentzian fit reveals a Q of 12 million. To ensure the validity of this result, the measurement is done at the undercoupled regime, with low power and in a region where the resonance is thermally broaden [22] and not narrowed.

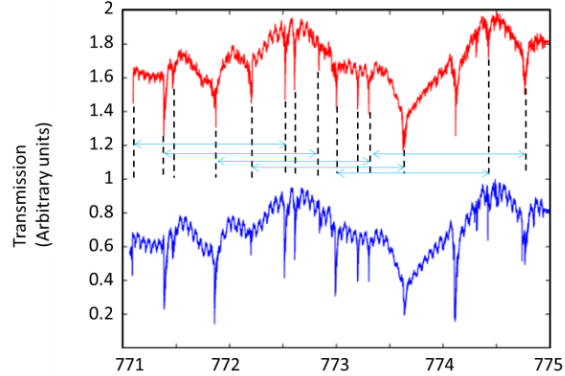

Fig .3. **Experimental results: resonances spectrum** exhibits several resonances at a 4 nm scan. To check that these resonances are repeatable, we scan while increasing wavelength (red) and then while decreasing wavelength (blue) and verify that resonances indeed stay. The blue arrows indicate the free spectral range.

We will now move to a broader scan that contains several resonances such as the one shown in figure 3. As one can see in figure 3, scanning along a 4 nm band reveals approximitly 16 resonances as expected for such spherical resonators. These resonances appear while scanning the wavelength up (Fig. 3 red) as well as while scanning it down (Fig. 3 blue). Such a dense spectrum can serve in future for phase matching for a variety of nonlinear effects including four waves mixing[23], third-harmonic generation[24], as well as backward[25, 26] and forward[27, 28] Brillioun scattering. An interesting detail of the spectra shown in Fig. 3 is that many of the peaks have a "sibling" located 1.4 nm away (blue arrows). Accordingly, in this regard, each of the cavity modes is expected to have an adjacent mode for which the number of optical waves along circumference is different by 1. The distance between these adjacent modes is generally referred to as the free spectral range, $FSR = \dfrac{\lambda^2}{2\pi R \cdot n_{eff}}$ where lambda is the vacuum wavelength, R is the resonator radius and $n_{eff}$ is the effective refractive index of the mode. As expected, the cavity radius calculated from the free spectral range (R=88μm) is close to the droplet radius measured with our microscope (R=85μm). Experimental error of 3% here comes from the resolution limit of the microscope, from uncertainties in refractive index and relates with the fact that the tunability of our laser is not perfectly linear with its control input.

We will now control the coupling distance while monitoring transmission, and demonstrate transition from the undercoupled regime to the critical one[29] when transmission drops to nearly zero[20].

Firstly, by moving the optical trap,we bring the resonator closer to the taper, and stop just as the absorption peak shows itself (figure 4a). The transmission at resonance drops by 10%, indicating being in the undercoupled regime. We then bring the sphere about ¼ µm closer to the taper and see that transmission drops by 93%, indicating a nearly critical coupling. Interestingly, beyond a certain point the optical trap cannot push the drop any closer to the taper. Since interfacial tension between the taper and water is large compared to taper and oil, a small gap between the taper and the resonator is always kept. If interfacial tension was not as is, then our experiment could end by a catastrophic collapse of the droplet towards coating the taper. Repulsion between drop and taper is therfore benefiting the durability of our resonator.

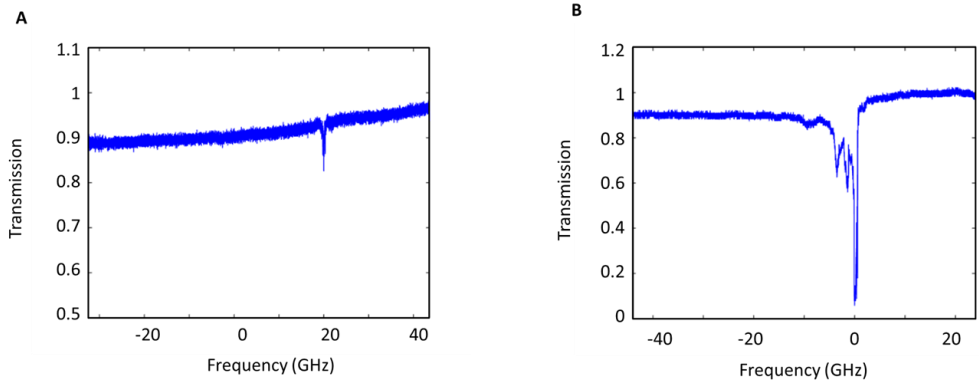

Fig. 4. **Experimental results, tunable coupling** from the undercoupled regime (A) to the critically coupled regime (B). During this measurement, the resonator is first trapped relatively far from the taper (A) and then brought closer to it (B).

In conclusion, we prove the concept of using an optical trap for activating oil droplets as fiber-coupled microresonators. We believe that our technique will soon extend to several resonators and then to an optical circuit where the shape and position of many optical devices will be controlled. Our technique might enable an adds-on when compared to lithography, by being able to test the resonators during fabrication and positioning them (as we demonstrate in figure 4). When such an optical inspection will reveal that the optical circuit is properly functioning, curing techniques will be used to turn the liquid circuit into a durable solid one. Though many challenges are still on our way, optical traps can wave major technology stoppers in manipulating multiple optical devices while precisely controling their position.